\documentclass{PoS}

\usepackage{xspace}
\newcommand{\GeVc}{\ensuremath{\mbox{Ge\kern-0.1em V}\!/\!c}\xspace}
\newcommand{\NASixtyOne}{NA61\slash SHINE\xspace}
\newcommand{\dedx}{\ensuremath{{\rm d}E\!/\!{\rm d}x}\xspace}
\newcommand{\Epos}{{\scshape Epos}\xspace}

 \title{Two-particle correlations in azimuthal angle and pseudorapidity in Be+Be} 
 \ShortTitle{Two-particle correlations in azimuthal angle and pseudorapidity in Be+Be}

 \author{\speaker{Bartosz Maksiak} for the \NASixtyOne Collaboration\\
        Warsaw University of Technology, Poland\\
        E-mail: \email{maksiak@if.pw.edu.pl}}

 \abstract{
 The \NASixtyOne experiment aims to discover the critical point of strongly interacting matter and study the properties of the onset of deconfinement. These goals are to be achieved by performing a two dimensional phase diagram $(T-\mu_B)$ scan by measurements of hadron production properties in proton-proton, proton-nucleus and nucleus-nucleus interactions as a function of collision energy and system size. In this contribution, the results on two-particle correlations in pseudorapidity and azimuthal angle will be presented for the first time for Be+Be interactions at beam momenta: 20, 30, 40, 75 and 150 GeV/c per nucleon. The \NASixtyOne results will be compared with the already presented results of proton-proton collisions at similar beam momenta as well as to the EPOS model results.
 }

 \FullConference{Critical Point and Onset of Deconfinement\\
          7-11 August, 2017\\
          The Wang Center, Stony Brook University, Stony Brook, NY}

\begin{document}

\section{Introduction}

Two-particle correlations in $\Delta\eta$, $\Delta\phi$ were studied extensively at RHIC and LHC. They allow to disentangle different sources of correlations: jets, flow, resonance decays, quantum statistics effects, conservation laws, etc. The \NASixtyOne experiment joined those studies as well and published results on two-particle correlations in p+p collisions at SPS energies~\cite{Maksiak:2015rxa,Aduszkiewicz:2016mww}. Those results showed structures that can be connected with resonance decays, momentum conservation and Bose-Einstein correlations.

This paper presents a continuation of studies on two-particle correlations in azimuthal angle and pseudorapidity. Its main purpose is to discover correlation structures and their possible sources in Be+Be collisions as well as to look for th differences between correlations in Be+Be and already published p+p results.

\section{$\Delta\eta\Delta\phi$ correlations}

Correlations are calculated as a function of the difference in pseudorapidity
($\eta$) and azimuthal angle ($\phi$) between two particles in the same event.
\begin{center}
  \begin{math}
    \Delta\eta = |{\eta}_1 - {\eta}_2|,
    \hspace{2cm}
    \Delta\phi = |{\phi}_1 - {\phi}_2|.
  \end{math}
\end{center}
The uncorrected ($raw$) correlation function is calculated as:
\begin{equation}
  \label{eq:correlations}
  C^{raw}(\Delta\eta,\Delta\phi)=
  \frac{N_{bkg}^{pairs}}{N_{signal}^{pairs}}
  \frac{S(\Delta\eta,\Delta\phi)}{B(\Delta\eta,\Delta\phi)},
\end{equation}
where
\begin{center}
  $S(\Delta\eta,\Delta\phi)=\frac{d^2N^{signal}}{d \Delta \eta d
    \Delta \phi}$; \hspace{0.1cm}
  $B(\Delta\eta,\Delta\phi)=\frac{d^2N^{bkg}}{d \Delta \eta d \Delta
    \phi}$
\end{center}
are the distributions for signal and background data, respectively and measured in restricted region $0 < \Delta\eta < 3$.
The $\Delta\phi$ range is folded, i.e. for $\Delta\phi$ larger than $\pi$ its value is recalculated as $2\pi - \Delta\phi$. In order to allow for a comparison with the RHIC and LHC results the pseudorapidity was calculated in the centre-of-mass (c.m.) system. The transformation from the laboratory system to the c.m. system was performed assuming the pion mass for all produced particles. Measured data was mirrored around $(\Delta\eta,\Delta\phi)=(0,0)$ point. Electrons and positrons were removed by a cut on \dedx (the energy loss of the particle tracks in the TPC detectors). Only the 5\% of the most violent Be+Be collisions were taken into account. The results from Be+Be collisions are not corrected for detector inefficiencies. The correction studies are currently ongoing.

\section{Results}

\begin{figure}
  \centering
  \includegraphics[width=0.32\textwidth]{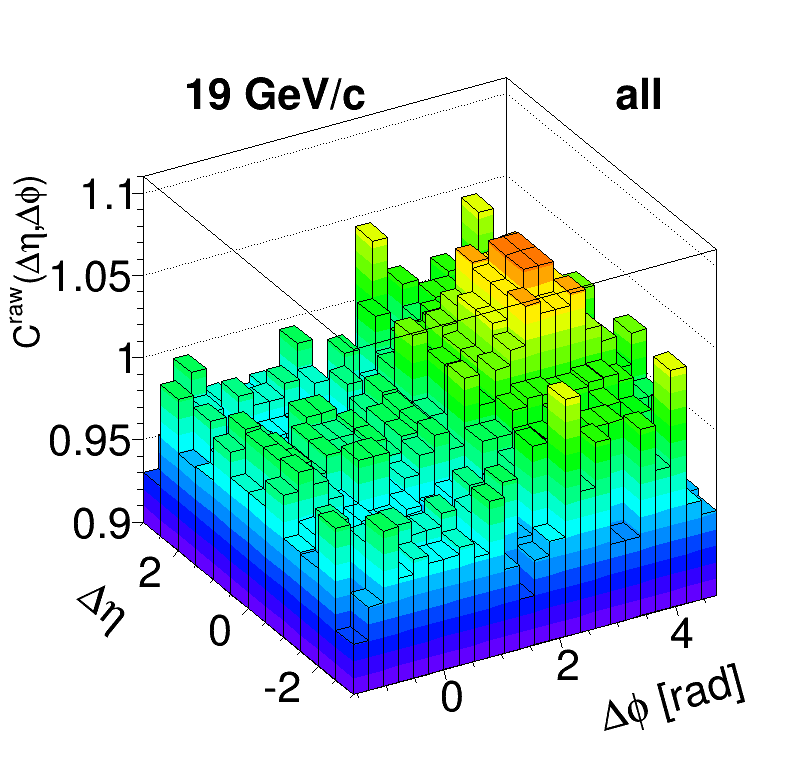}
  \includegraphics[width=0.32\textwidth]{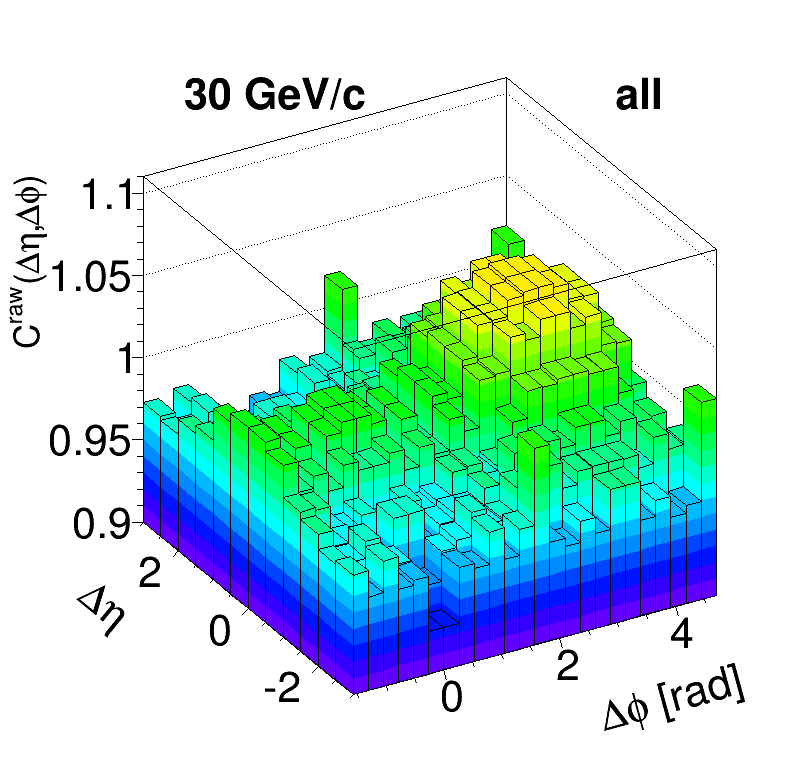}
  \includegraphics[width=0.32\textwidth]{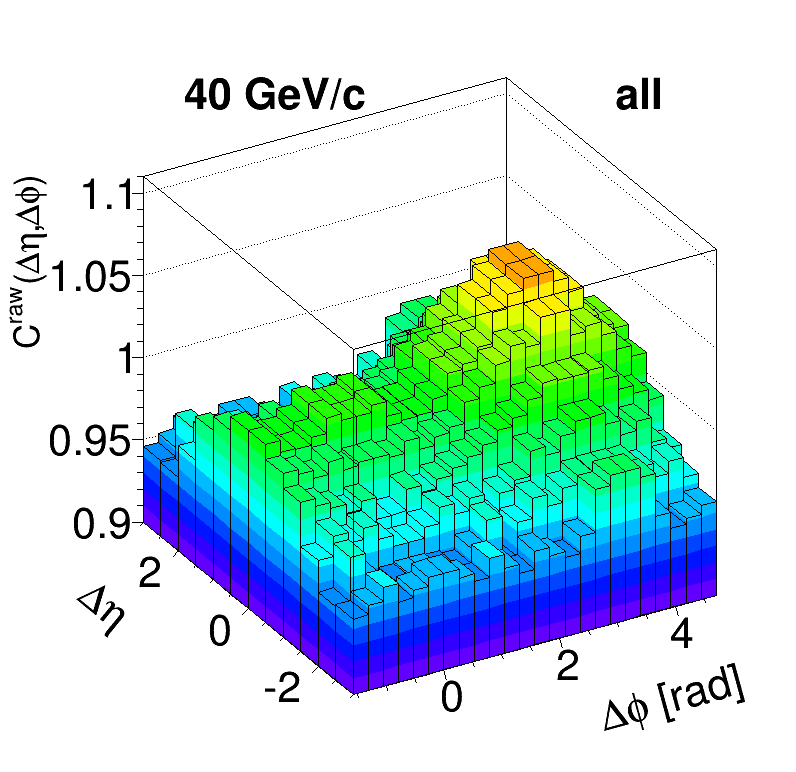}
  \\
  \includegraphics[width=0.32\textwidth]{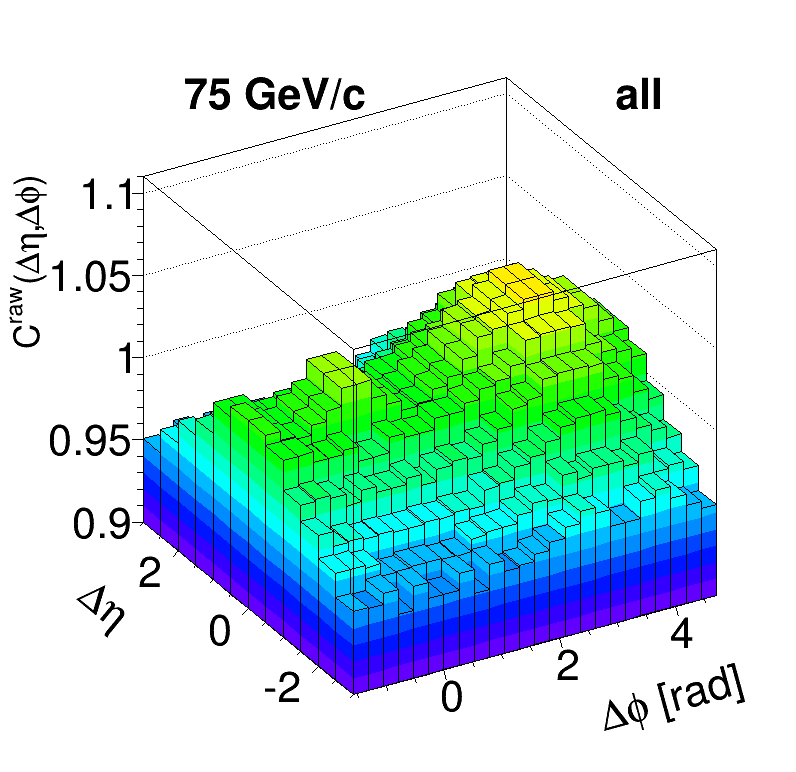}
  \includegraphics[width=0.32\textwidth]{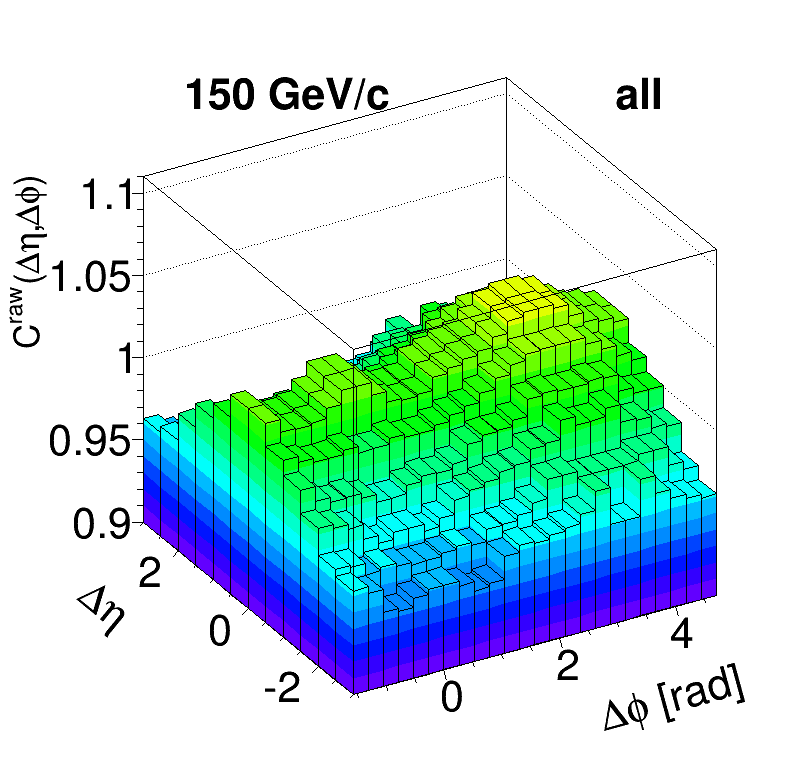}
  \caption{Results on two-particle correlations in $\Delta\eta\Delta\phi$ in Be+Be collisions presented for five beam momenta. All charged pairs of particles.}
  \label{fig:detadphi_BeBe_all}
\end{figure}

\begin{figure}
  \centering
  \includegraphics[width=0.32\textwidth]{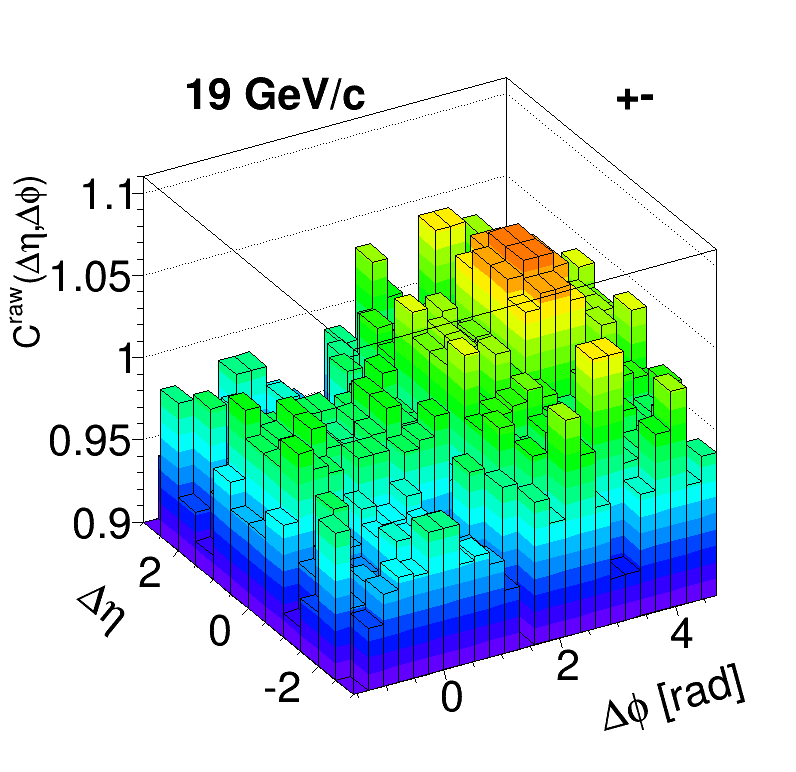}
  \includegraphics[width=0.32\textwidth]{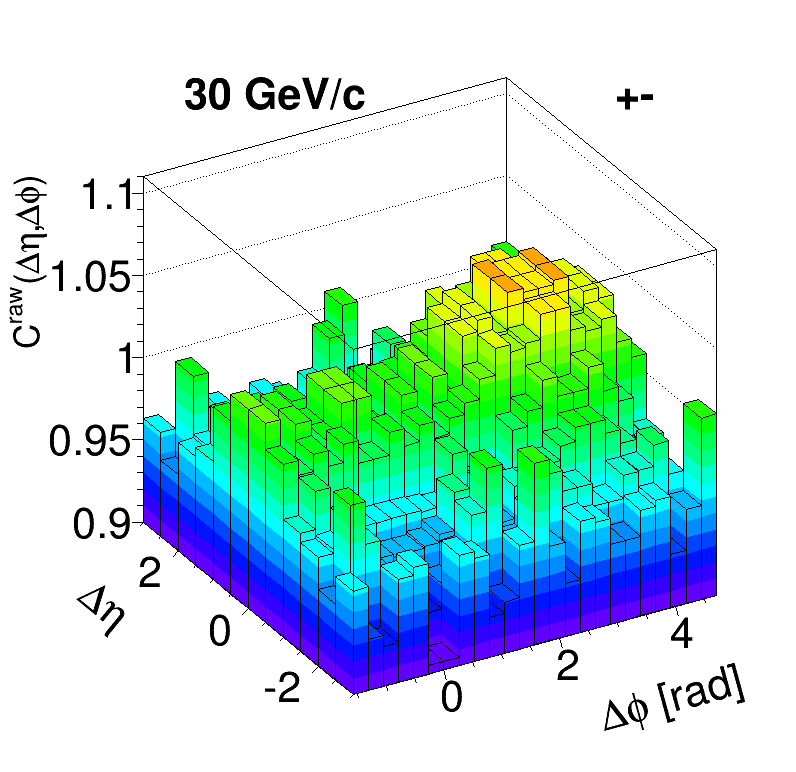}
  \includegraphics[width=0.32\textwidth]{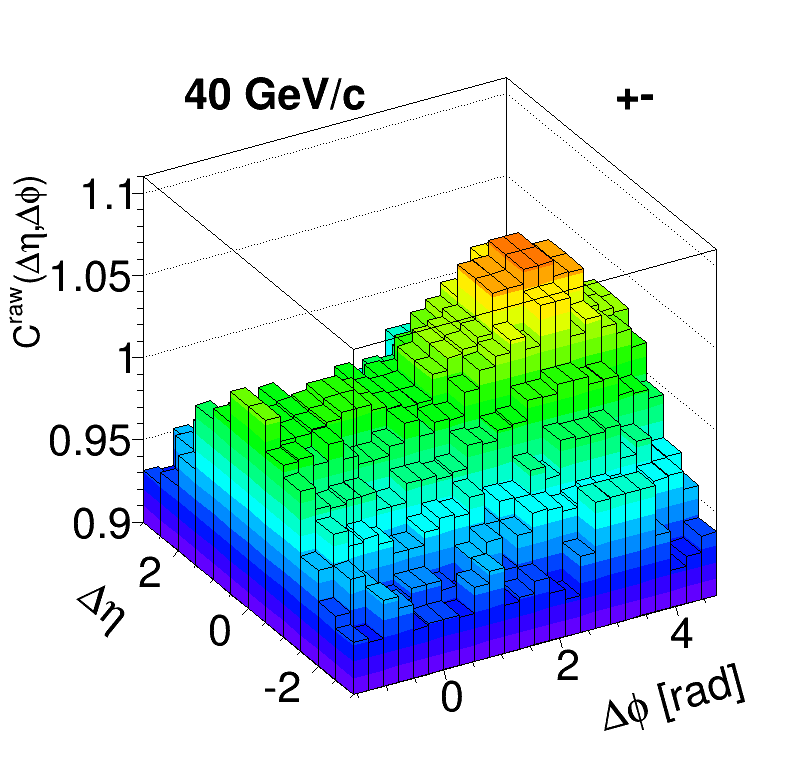}
  \\
  \includegraphics[width=0.32\textwidth]{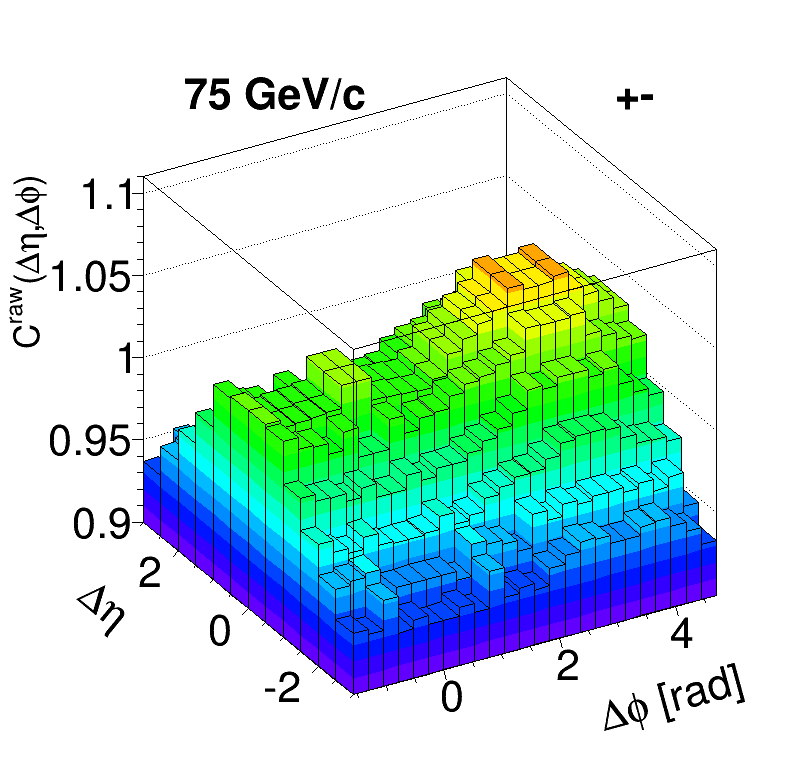}
  \includegraphics[width=0.32\textwidth]{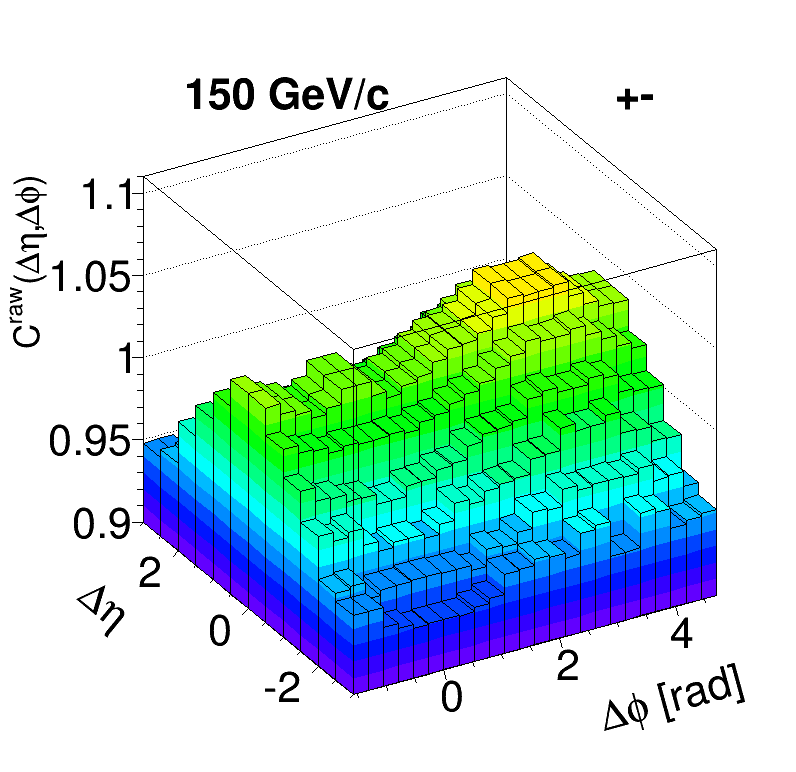}
  \caption{Results on two-particle correlations in $\Delta\eta\Delta\phi$ in Be+Be collisions presented for five beam momenta. Unlike-sign pairs of particles.}
  \label{fig:detadphi_BeBe_unlike}
\end{figure}

\begin{figure}
  \centering
  \includegraphics[width=0.32\textwidth]{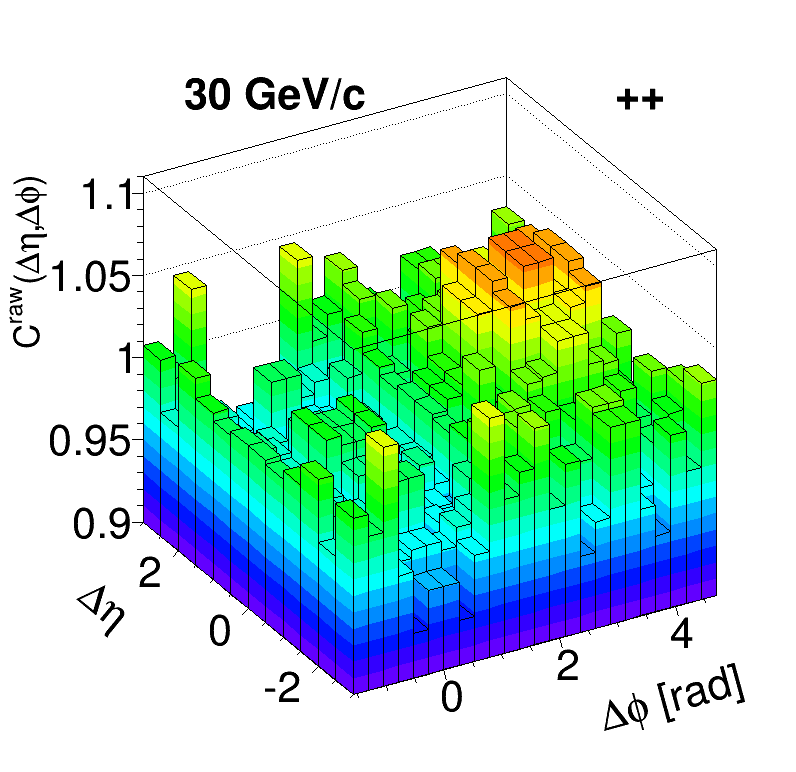}
  \includegraphics[width=0.32\textwidth]{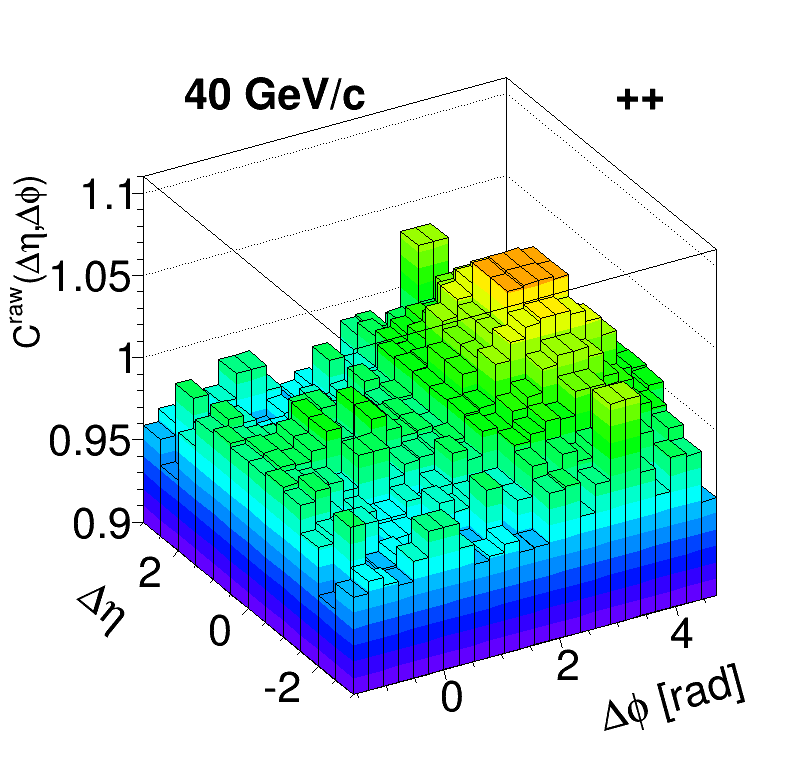}
  \\
  \includegraphics[width=0.32\textwidth]{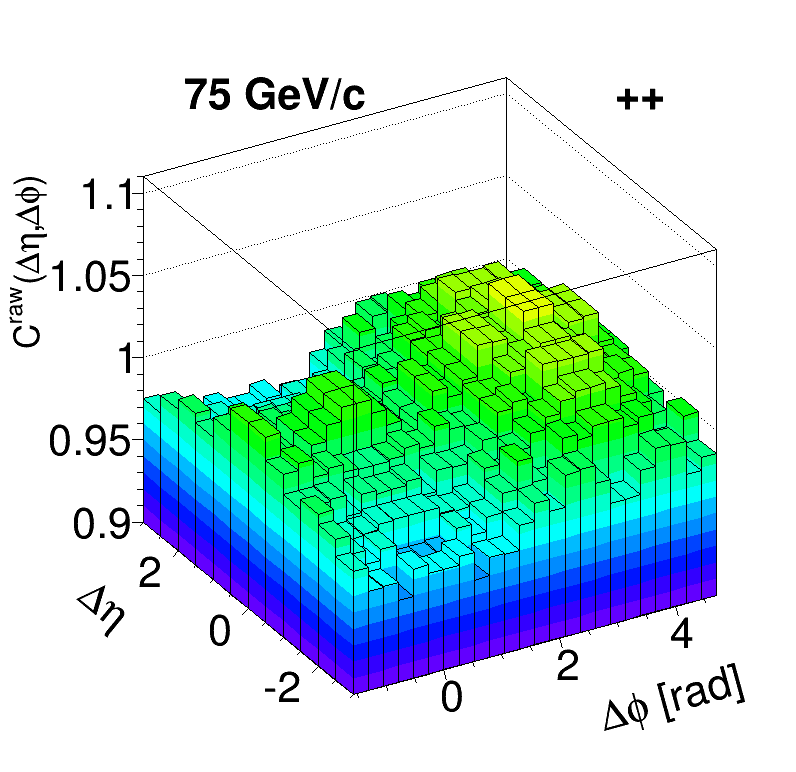}
  \includegraphics[width=0.32\textwidth]{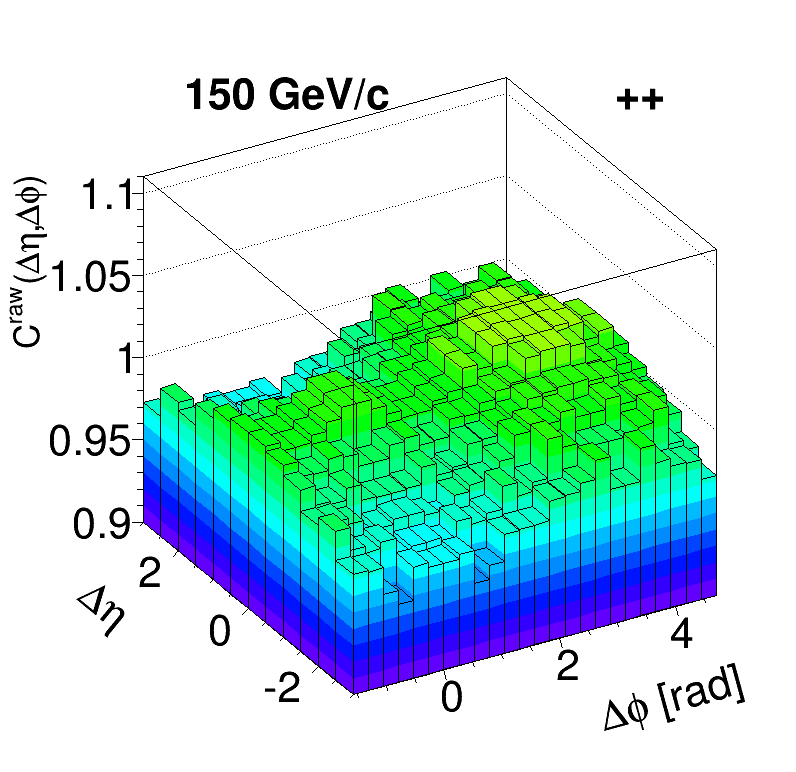}
  \caption{Results on two-particle correlations in $\Delta\eta\Delta\phi$ in Be+Be collisions presented for four beam momenta (19$A$~GeV/c was omitted due to low statistics). Positively charged pairs of particles.}
  \label{fig:detadphi_BeBe_pos}
\end{figure}

\begin{figure}
  \centering
  \includegraphics[width=0.32\textwidth]{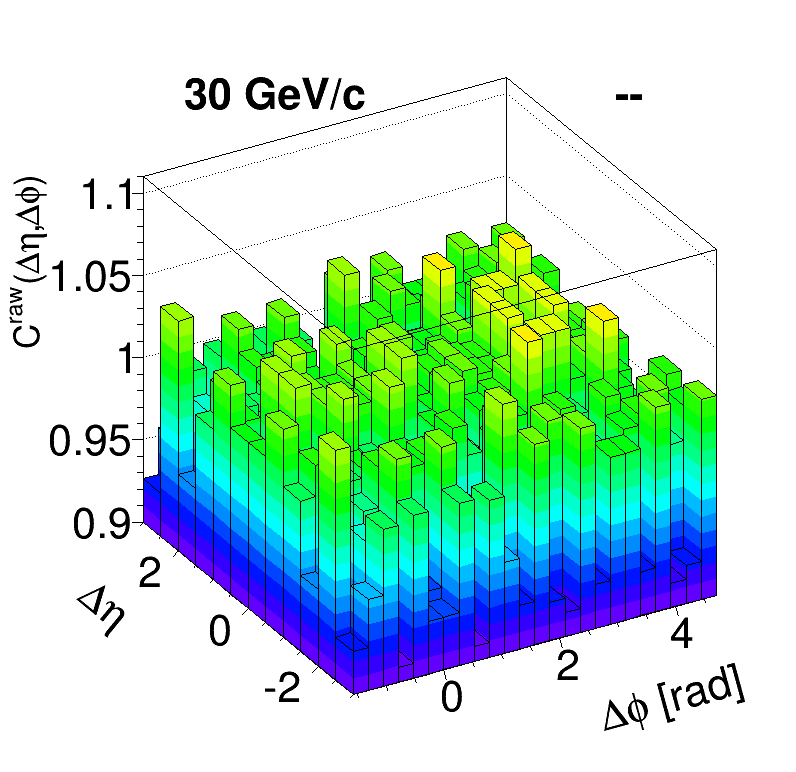}
  \includegraphics[width=0.32\textwidth]{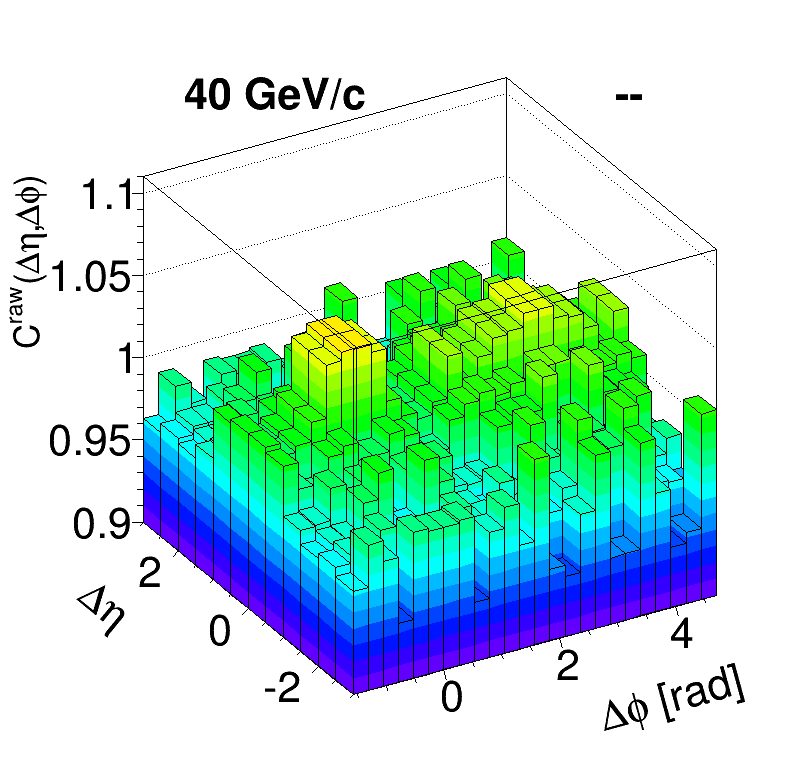}
  \\
  \includegraphics[width=0.32\textwidth]{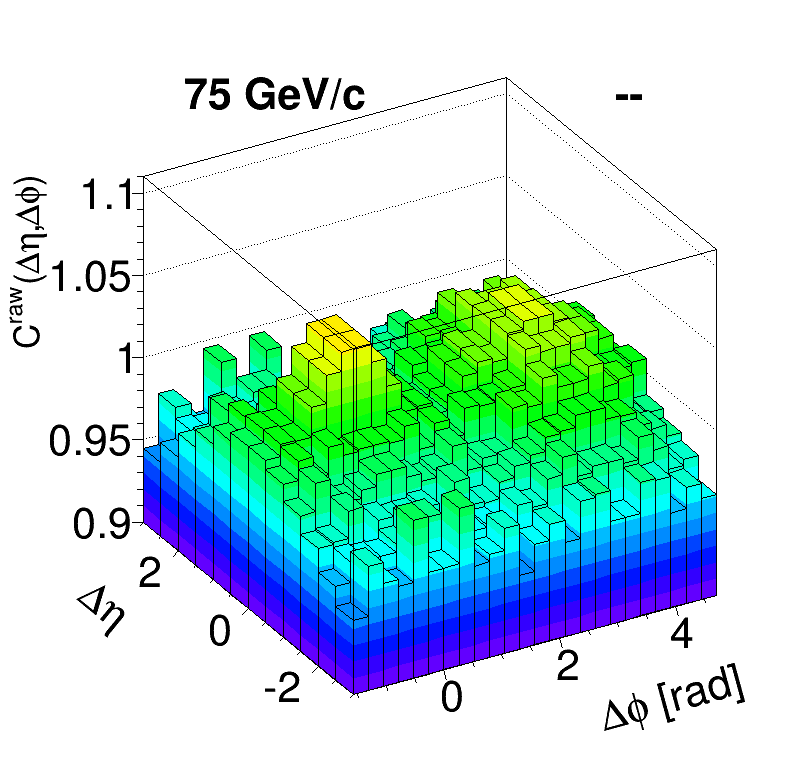}
  \includegraphics[width=0.32\textwidth]{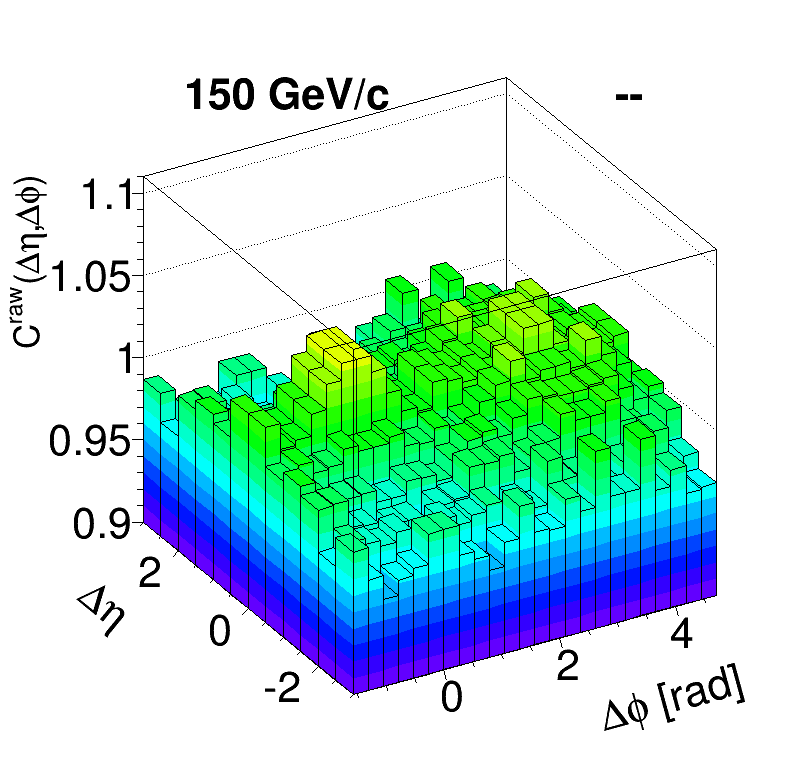}
  \caption{Results on two-particle correlations in $\Delta\eta\Delta\phi$ in Be+Be collisions presented for four beam momenta (19$A$~GeV/c was omitted due to low statistics). Negatively charged pairs of particles.}
  \label{fig:detadphi_BeBe_neg}
\end{figure}

This section presents preliminary results from the analysis of two-particle correlations in azimuthal angle and pseudorapidity in Be+Be collisions. Figures \ref{fig:detadphi_BeBe_all}, \ref{fig:detadphi_BeBe_unlike}, \ref{fig:detadphi_BeBe_pos}, and~\ref{fig:detadphi_BeBe_neg} show energy scan results for all charged, unlike-sign, positively charged and negatively charged pairs of particles, respectively.

There are two main structures visible in Be+Be results. The most prominent is a maximum at $(\Delta\eta,\Delta\phi) = (0,\pi)$ probably due to resonance decays and momentum conservation. The away-side enhancement is the highest in unlike-sign pairs and weaker in like-sign pairs. This indicates that the largest contribution to it comes from correlations of particles with opposite signs, suggesting that this is due to resonances decaying back to back into two unlikely-signed particles. In positively charged pairs this enhancement is lower due to lower number of resonances decaying into two positively charged particles (e.g.~$\Delta^{++}$), but in negatively charged pairs it is hardly visible because the multiplicity of double-negative resonances, decaying into two negatively charged particles, is very low.

Another, well-visible structure is a small maximum at $(0,0)$ which is present for all charge combinations. For unlike-sign pairs its most probable source is Coulomb attraction. In like-sign pairs it is probably due to quantum statistics. A difference in its height between positive and negative particle pairs can be observed. Its probable explanation is that in negatively charged pairs of particles the most frequently produced particles are negative pions, which are bosons and thus they obey Bose-Einstein statistics giving the enhancement. In positively charged pairs of particles however, the most abundantly produced particles are positively charged pions (bosons) and protons being fermions and following Fermi-Dirac statistics principles. Correlations for positively charged pairs of particles is an interplay of those two effects: enhancement for bosons and suppression for fermions, and therefore correlations for positively charged particles still produce enhancement but lower than in correlations of negatively charged pairs with no significant production of fermions.

\section{Comparisons}

In this section, results from real Be+Be data are compared to the \Epos model~\cite{Werner:2005jf} results and p+p results already published by the \NASixtyOne experiment~\cite{Aduszkiewicz:2016mww}.

\subsection{Comparison with EPOS}

\begin{figure}
  \centering
  \includegraphics[width=0.195\textwidth]{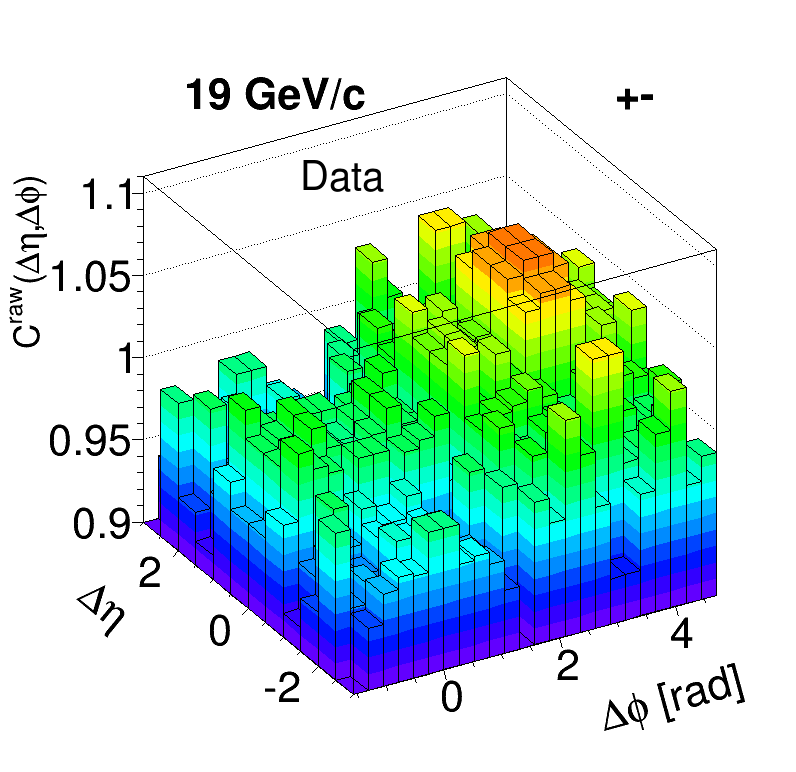}
  \includegraphics[width=0.195\textwidth]{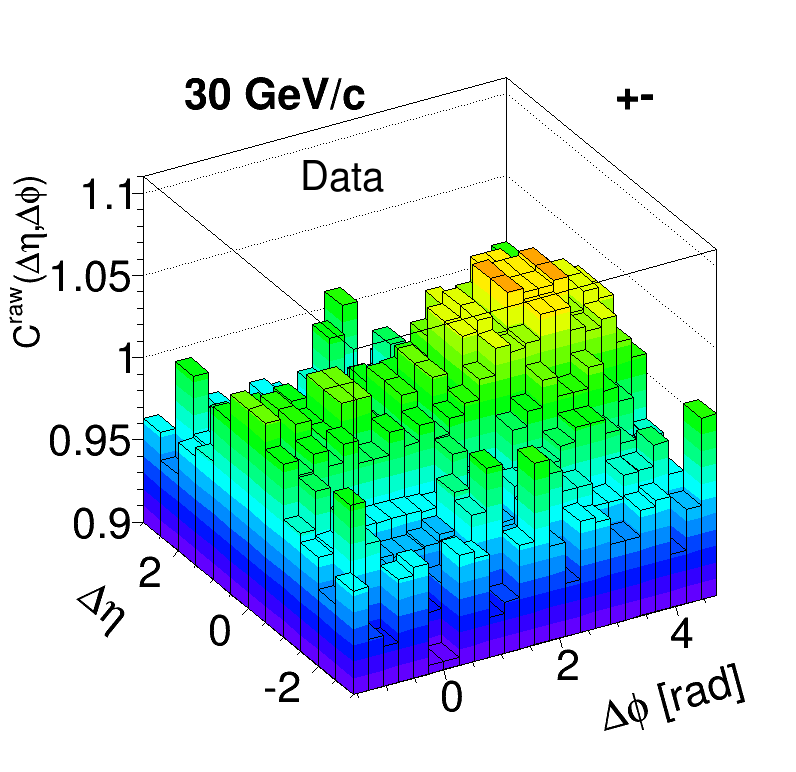}
  \includegraphics[width=0.195\textwidth]{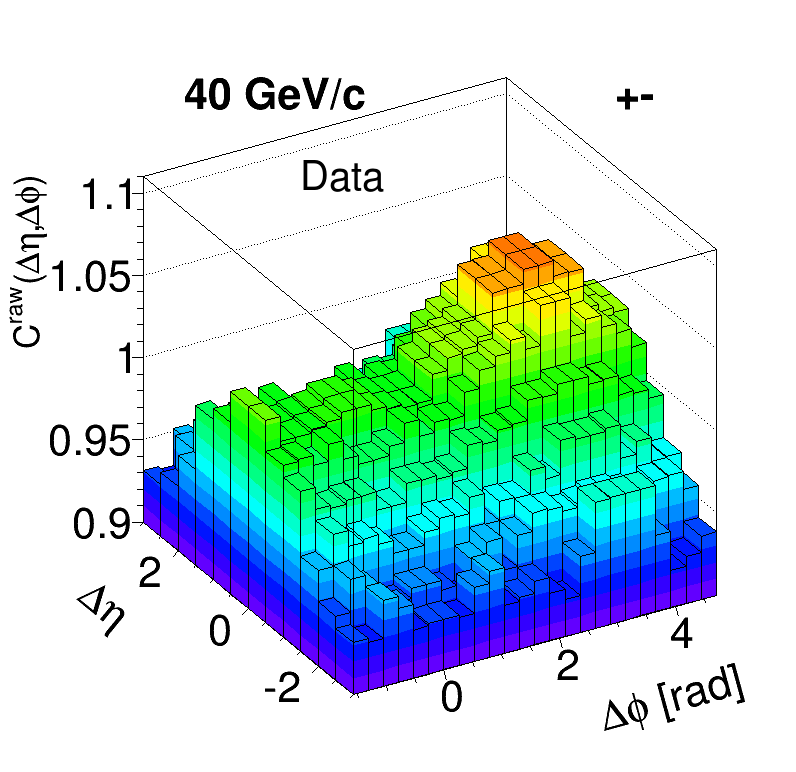}
  \includegraphics[width=0.195\textwidth]{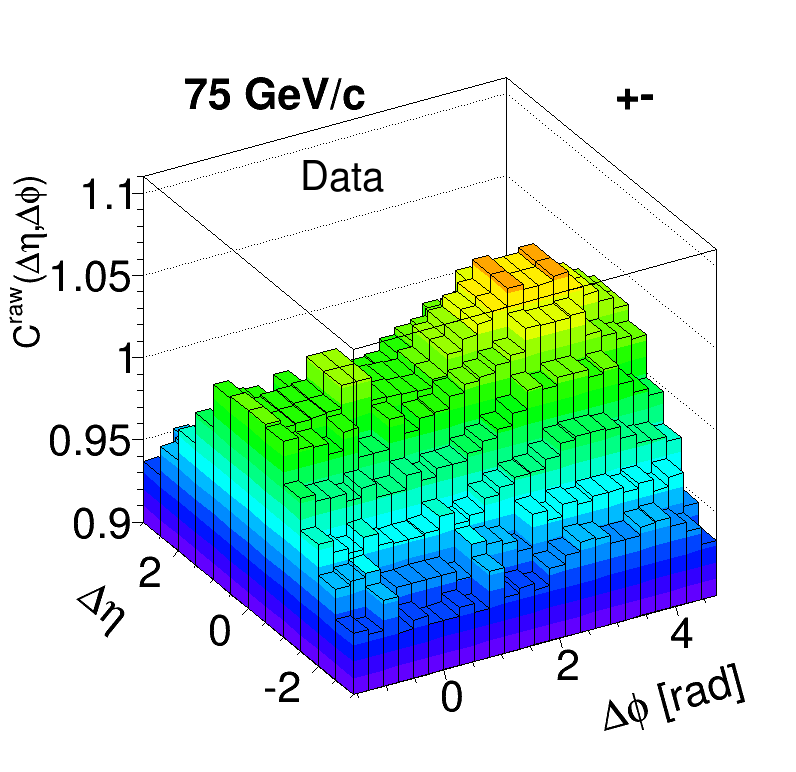}
  \includegraphics[width=0.195\textwidth]{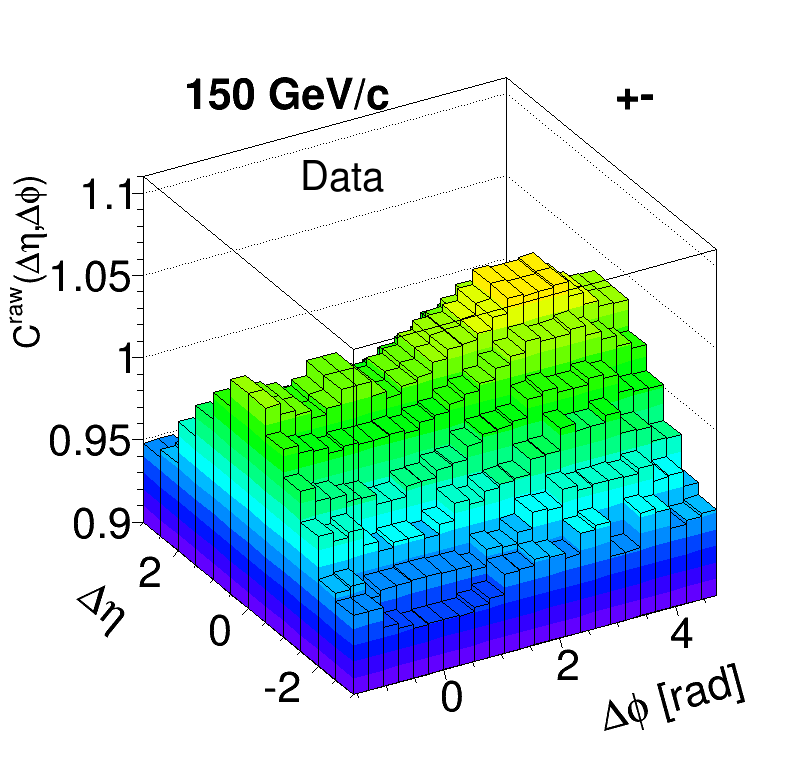}
  \\
  \includegraphics[width=0.195\textwidth]{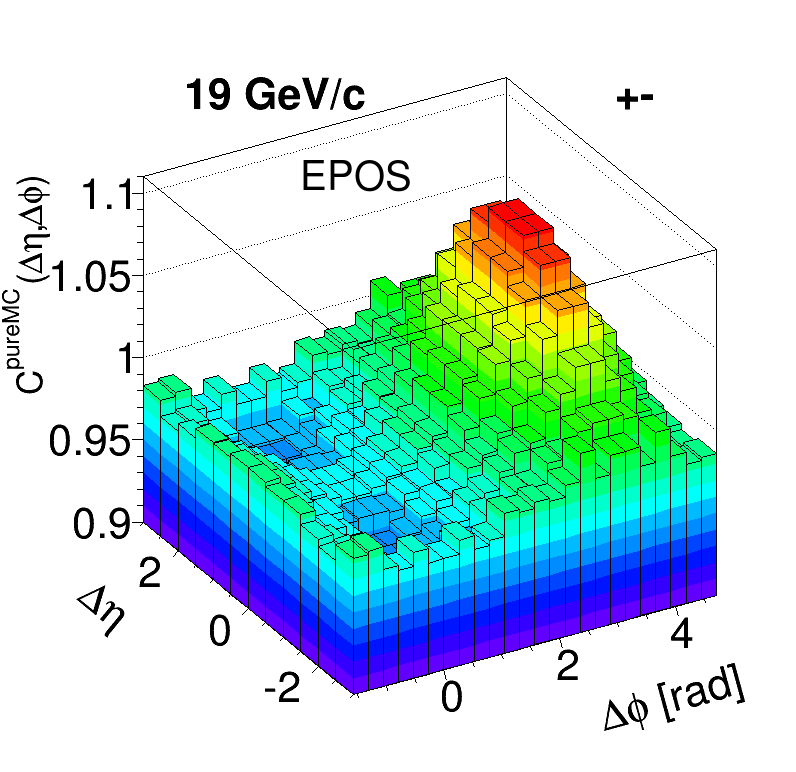}
  \includegraphics[width=0.195\textwidth]{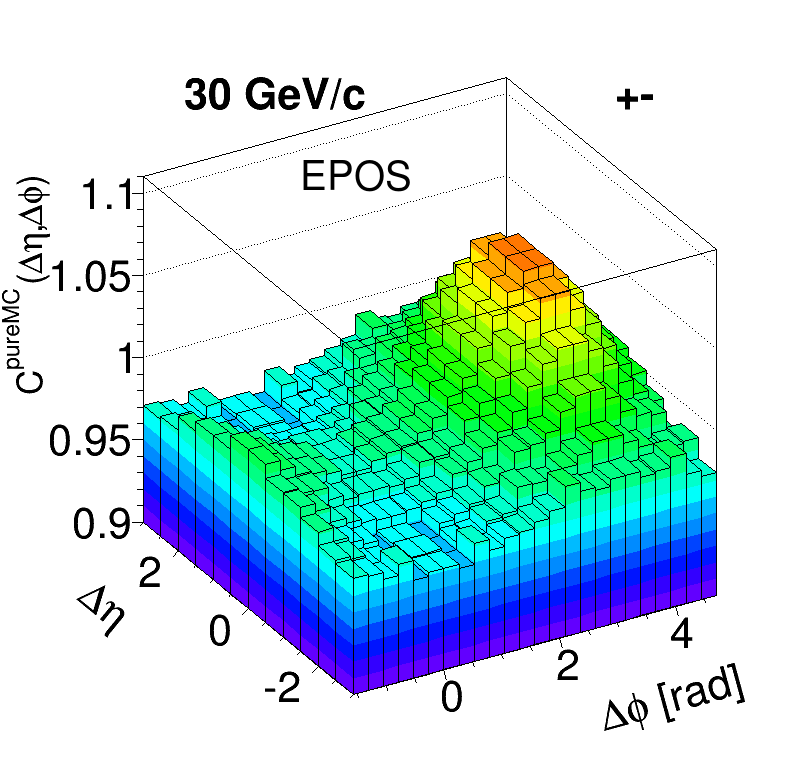}
  \includegraphics[width=0.195\textwidth]{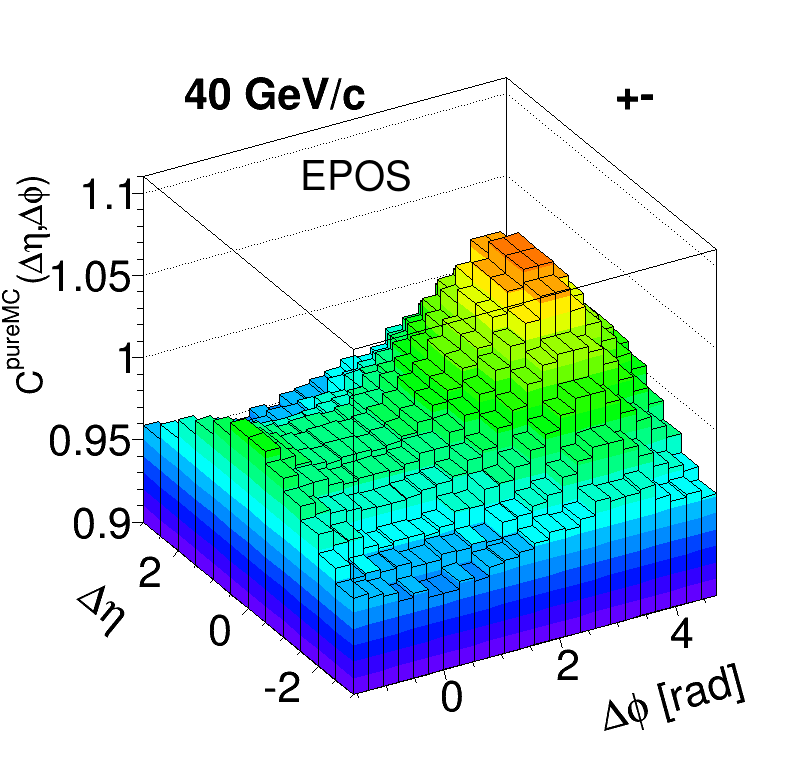}
  \includegraphics[width=0.195\textwidth]{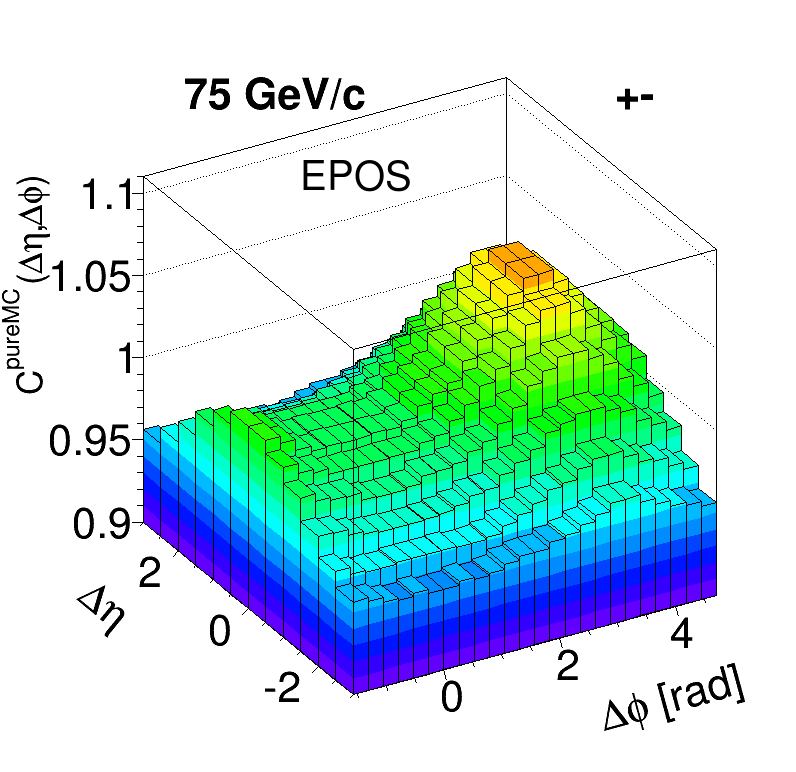}
  \includegraphics[width=0.195\textwidth]{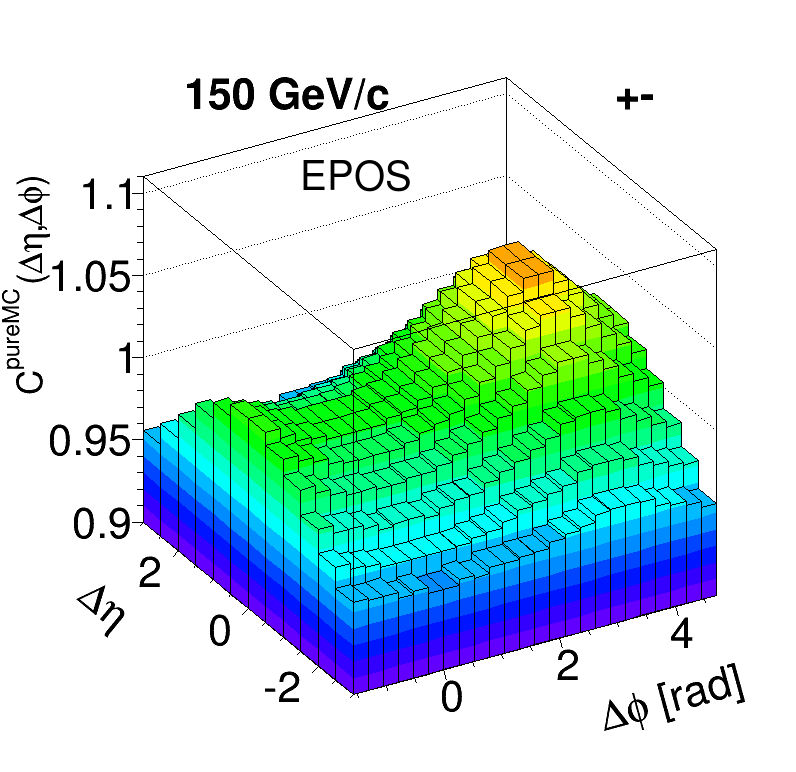}
  \caption{Comparison of Be+Be \NASixtyOne results (top row) and Be+Be results obtained in the EPOS model (bottom row). Results for unlike-sign pairs. EPOS results are within full acceptance.}
  \label{fig:Data_vs_EPOS_unlike}
\end{figure}

\begin{figure}
  \centering
  \includegraphics[width=0.24\textwidth]{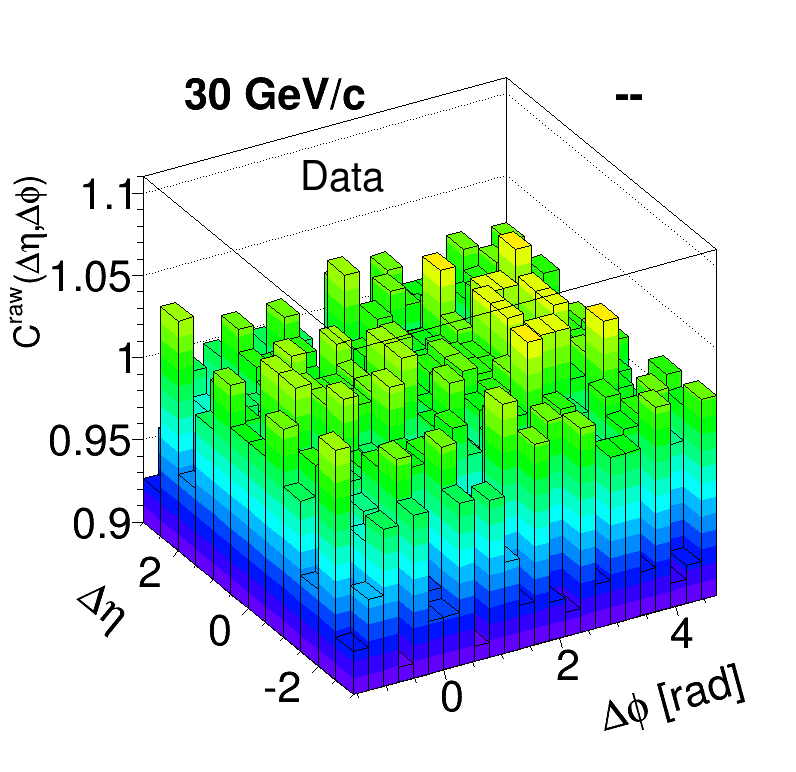}
  \includegraphics[width=0.24\textwidth]{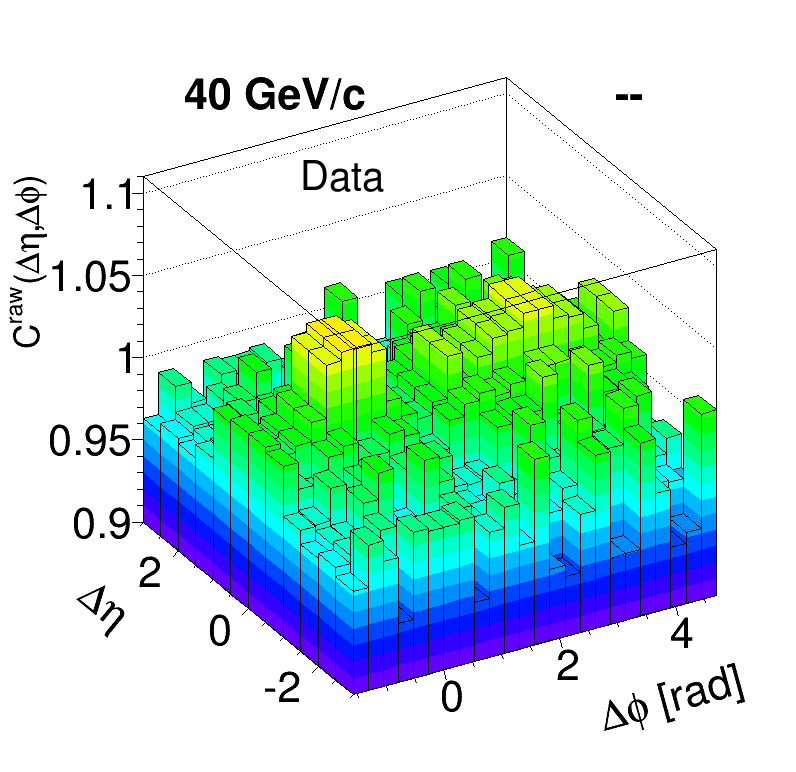}
  \includegraphics[width=0.24\textwidth]{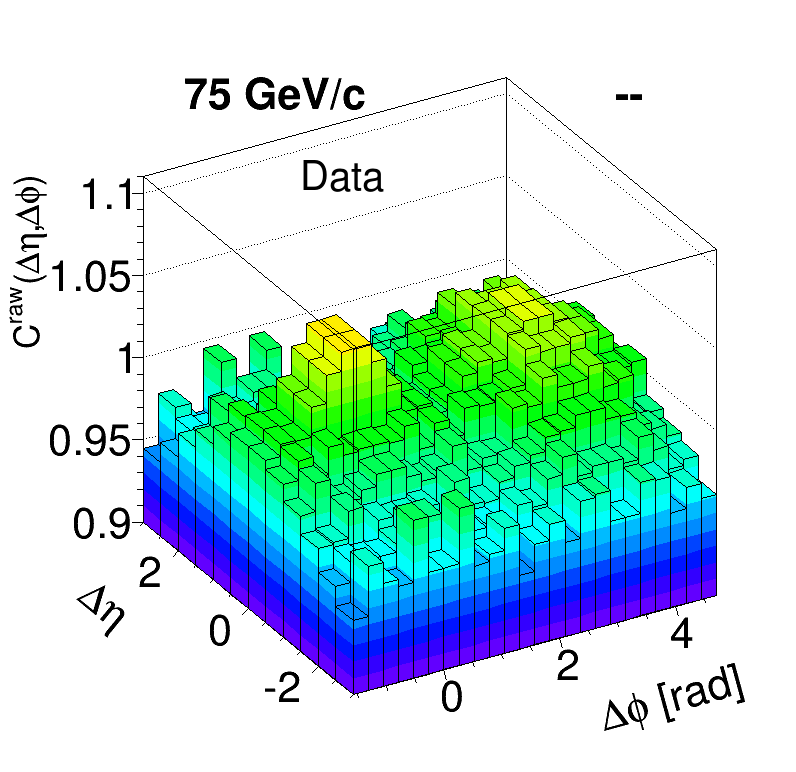}
  \includegraphics[width=0.24\textwidth]{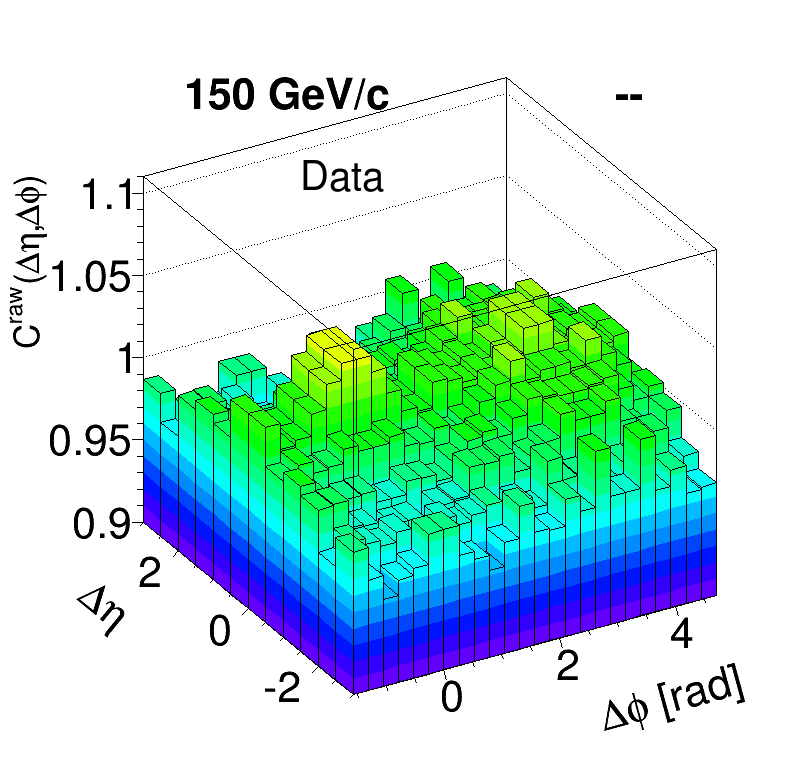}
  \\
  \includegraphics[width=0.24\textwidth]{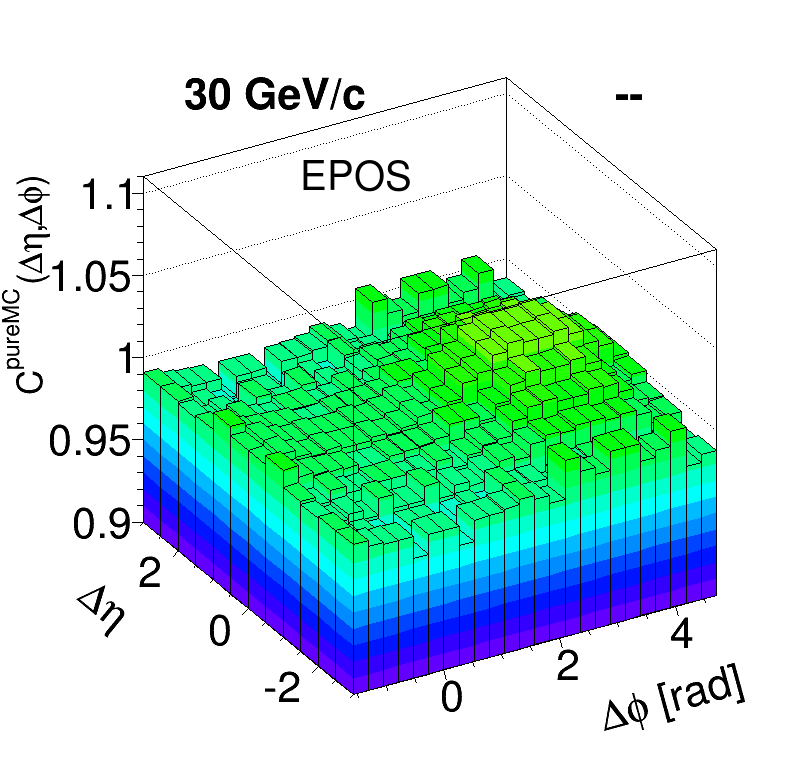}
  \includegraphics[width=0.24\textwidth]{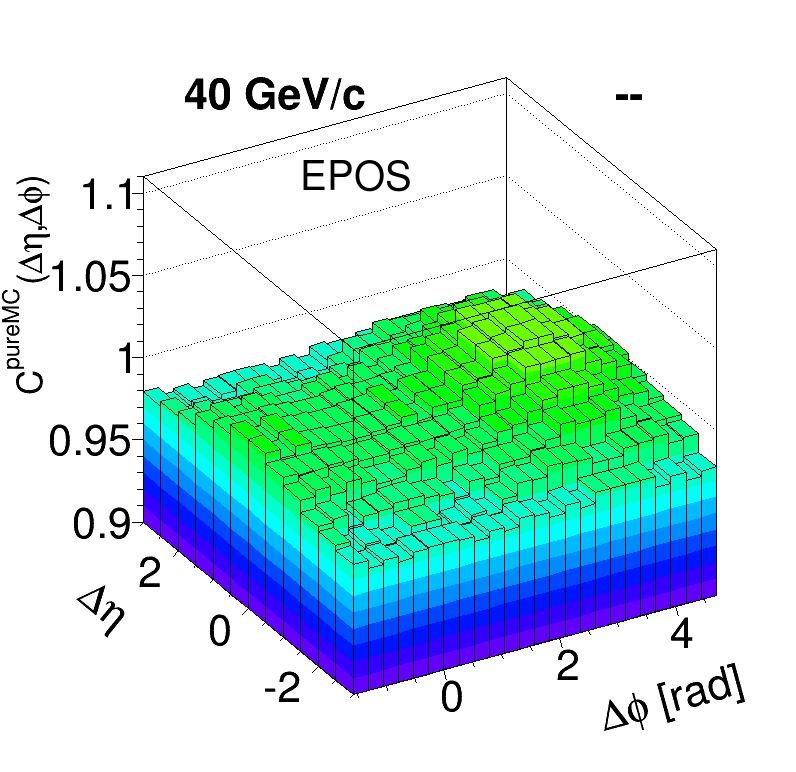}
  \includegraphics[width=0.24\textwidth]{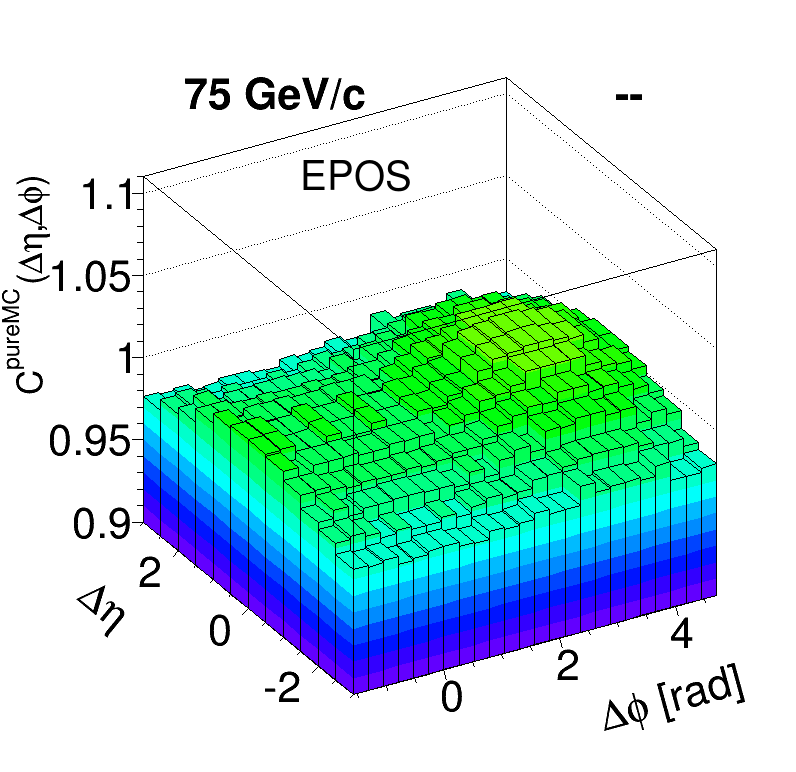}
  \includegraphics[width=0.24\textwidth]{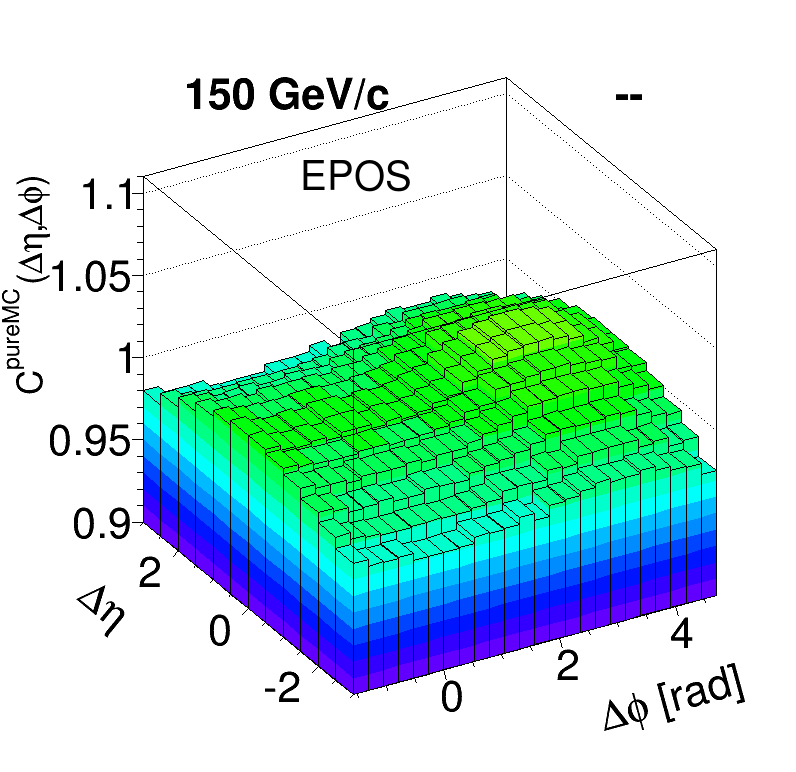}
  \caption{Comparison of Be+Be \NASixtyOne results (top row) and Be+Be results obtained in the EPOS model (bottom row). Results for negatively charged pairs of particles. EPOS results are within full acceptance.}
  \label{fig:Data_vs_EPOS_neg}
\end{figure}

\begin{figure}
  \centering
  \includegraphics[width=0.24\textwidth]{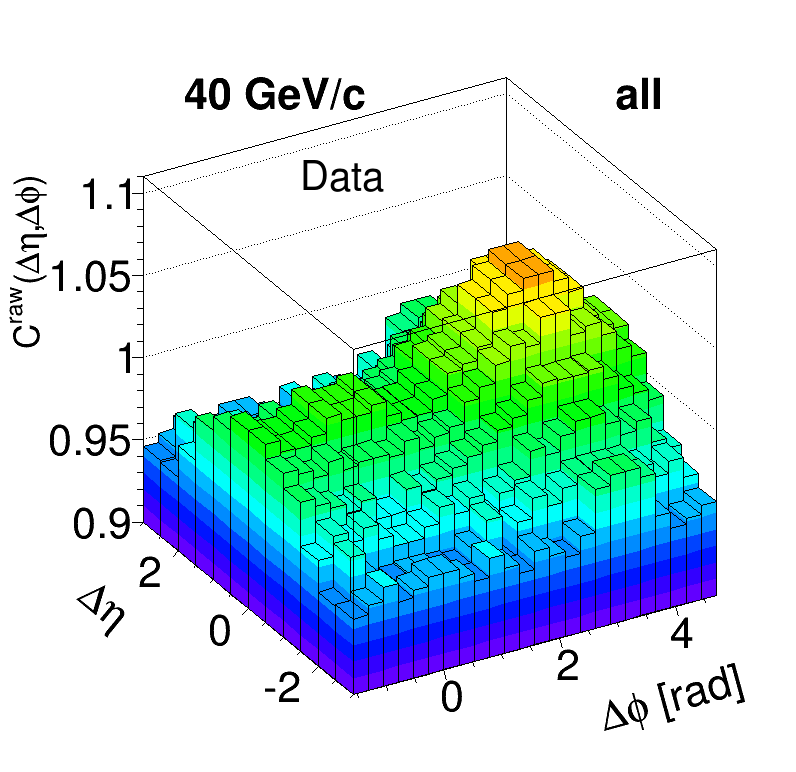}
  \includegraphics[width=0.24\textwidth]{Figs/DetaDphi/40/Correlations_alicestyle_unlike_text_data.png}
  \includegraphics[width=0.24\textwidth]{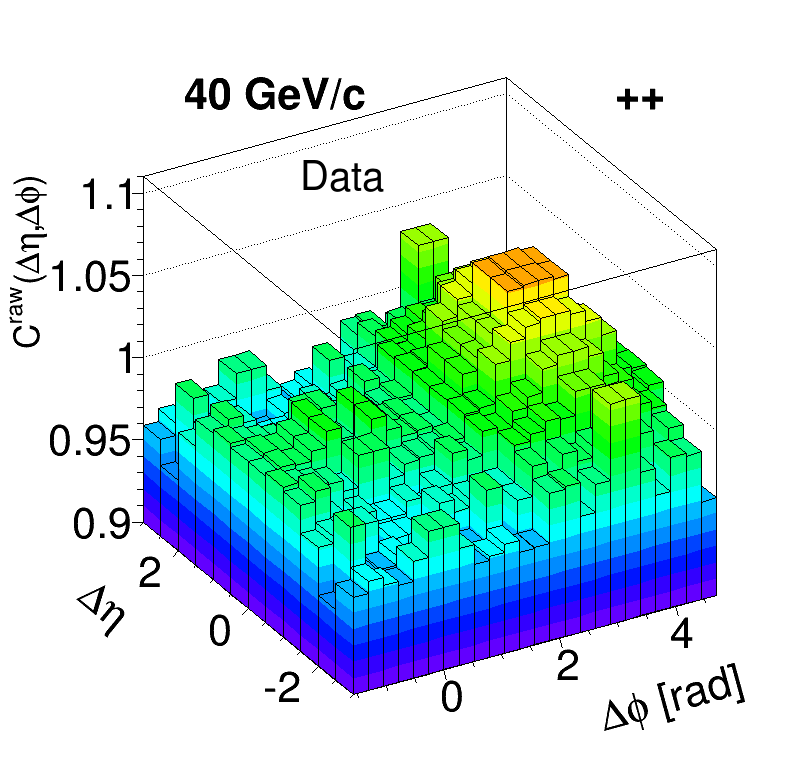}
  \includegraphics[width=0.24\textwidth]{Figs/DetaDphi/40/Correlations_alicestyle_neg_text_data.png}
  \\
  \includegraphics[width=0.24\textwidth]{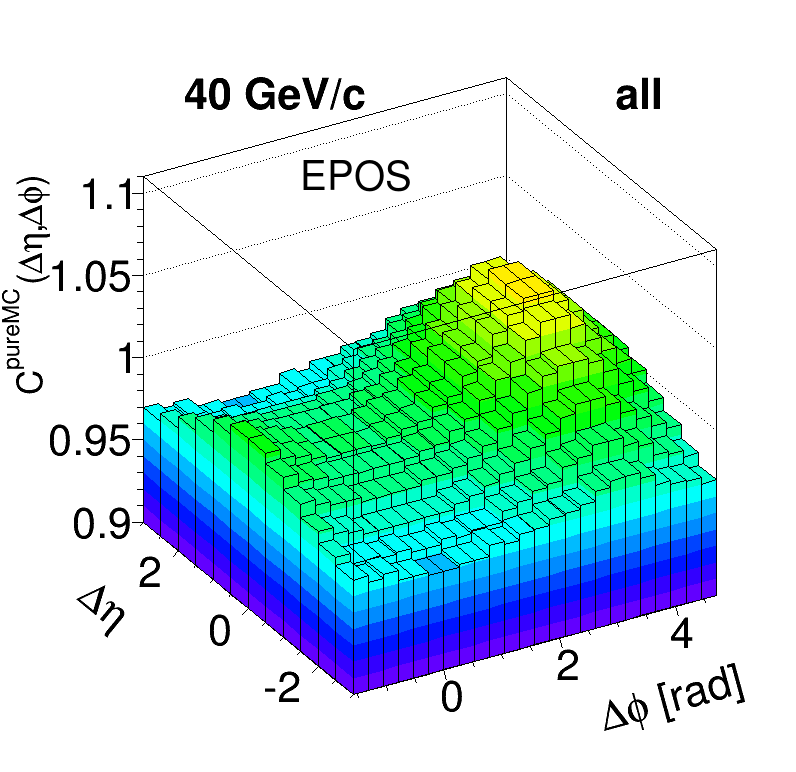}
  \includegraphics[width=0.24\textwidth]{Figs/EPOS_4pi/40/Correlations_alicestyle_unlike_text_epos.png}
  \includegraphics[width=0.24\textwidth]{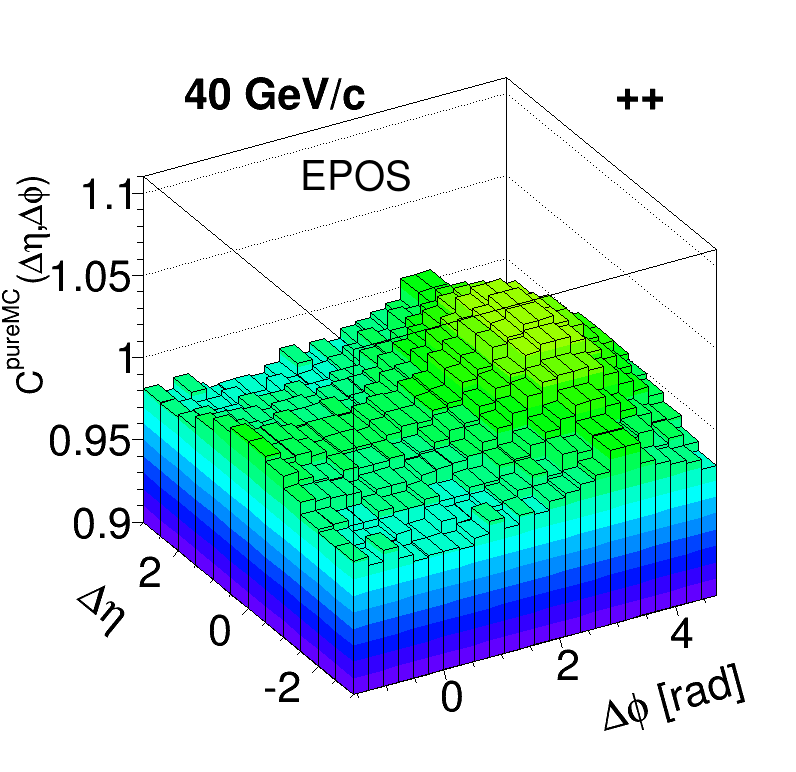}
  \includegraphics[width=0.24\textwidth]{Figs/EPOS_4pi/40/Correlations_alicestyle_neg_text_epos.png}
  \caption{Charge dependence comparison of Be+Be \NASixtyOne results (top row) and Be+Be data generated in the EPOS model (bottom row). Results presented for 40$A$~GeV/c beam momentum. EPOS results are within full acceptance.}
  \label{fig:Data_vs_EPOS_charge_dep}
\end{figure}

Be+Be results were compared to data generated in \Epos. Note that \Epos generated results are presented in full ($4\pi$) acceptance. The example comparison for unlike-sign and negatively charged pairs is presented in Figs.~\ref{fig:Data_vs_EPOS_unlike} and \ref{fig:Data_vs_EPOS_neg}, respectively. Additional comparison of charge dependence for selected beam momentum (40$A$~GeV/c) is shown in Fig.~\ref{fig:Data_vs_EPOS_charge_dep}.

\Epos reproduces data qualitatively well. Similarly to real data, the away-side enhancement is the most pronounced at lower beam momenta. The differences between charge combinations are also reproduced by the \Epos model. However, \Epos does not produce peak at (0,0). This is due to the lack of implementation of short-range correlations, like quantum statistics and Coulomb interactions, in the model. As a result, correlations in (0,0) region for \Epos generated data are almost flat.

\newpage

\subsection{Comparison with results from p+p}

Correlations from Be+Be collisions were compared to p+p results~\cite{Aduszkiewicz:2016mww}.

\begin{figure}
  \centering
  \includegraphics[width=0.32\textwidth]{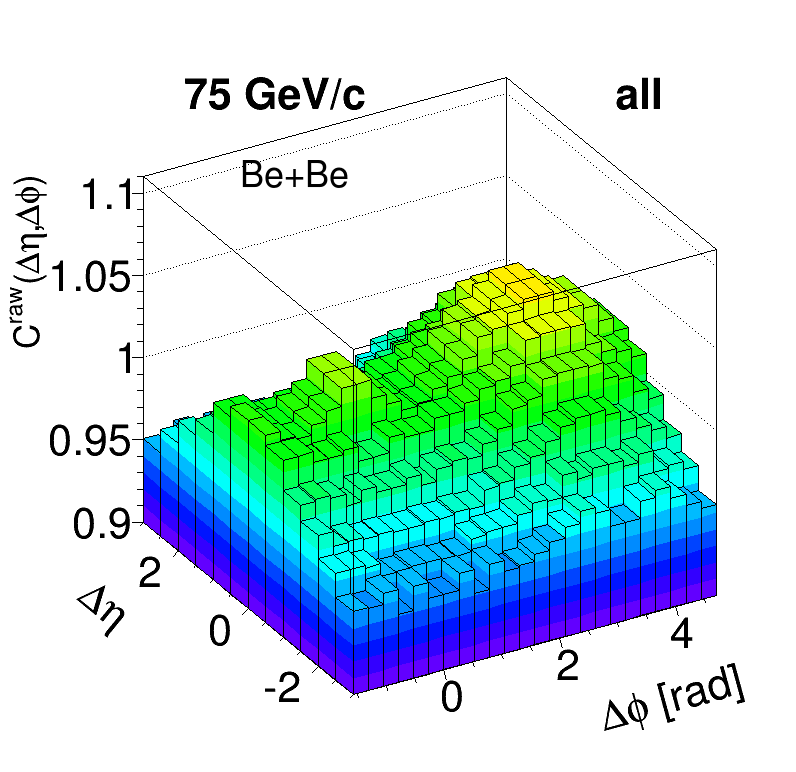}
  \includegraphics[width=0.32\textwidth]{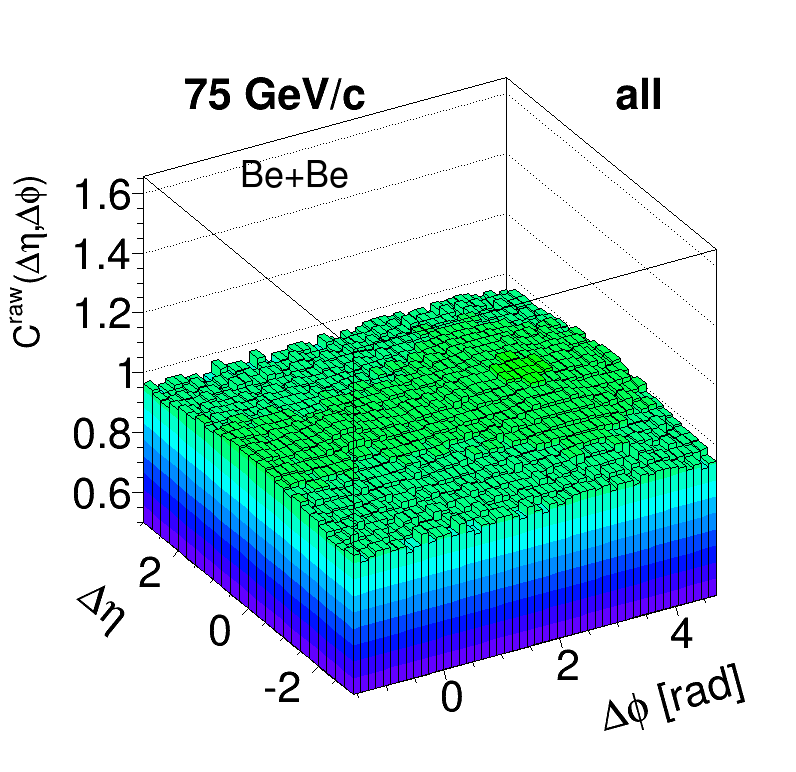}
  \includegraphics[width=0.32\textwidth]{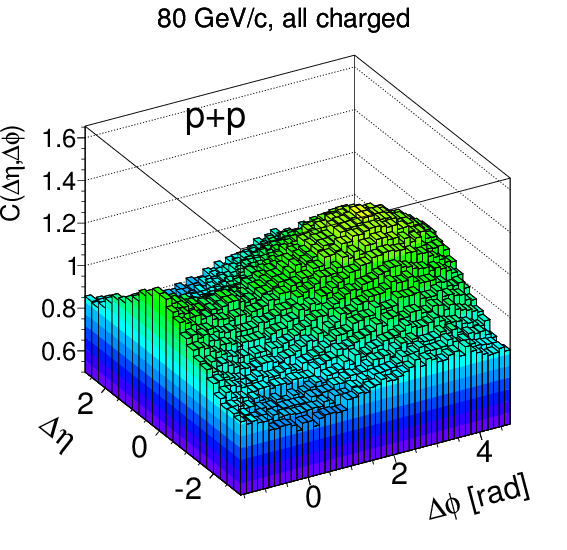}
  \caption{Comparison of correlations magnitude on an example of Be+Be at 75$A$~GeV/c and p+p~\cite{Aduszkiewicz:2016mww} at 80~GeV/c. Left and middle plots how the same data but in different vertical scales.}
  \label{fig:scales_comparison}
\end{figure}

It needs to be mentioned that correlations in Be+Be are generally much weaker than in p+p due to higher combinatorical background that dilutes the signal of correlations. The comparison of scales in Be+Be and p+p correlations is presented in Fig.~\ref{fig:scales_comparison}.

\begin{figure}
  \centering
  \includegraphics[width=0.24\textwidth]{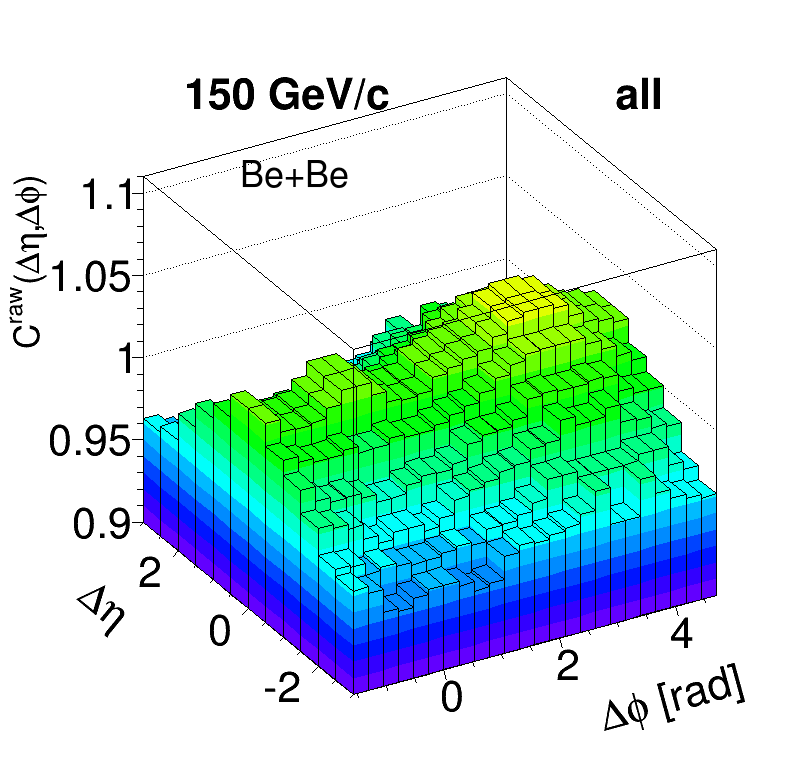}
  \includegraphics[width=0.24\textwidth]{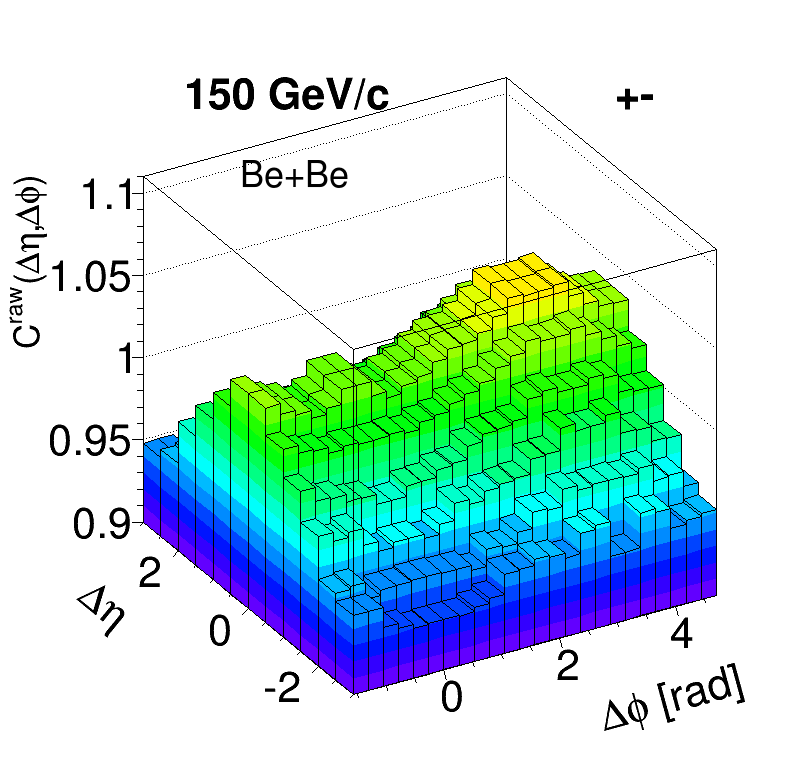}
  \includegraphics[width=0.24\textwidth]{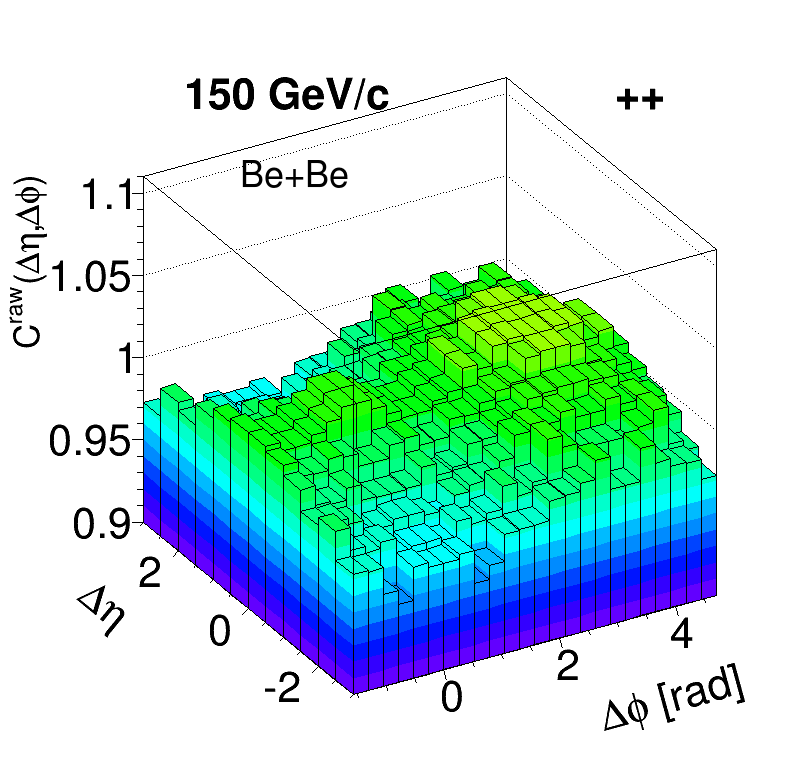}
  \includegraphics[width=0.24\textwidth]{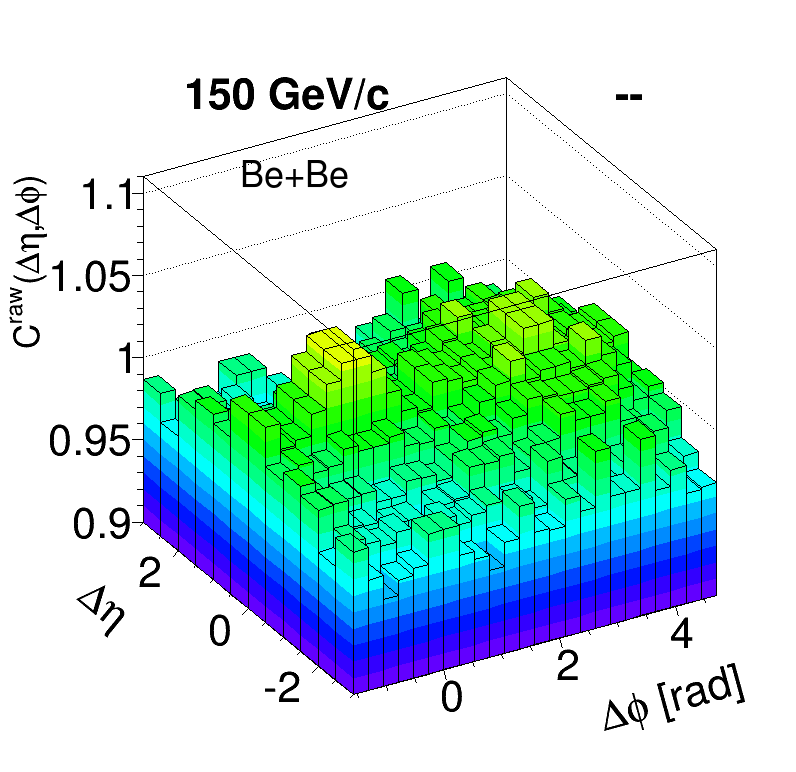}
  \\
  \includegraphics[width=0.24\textwidth]{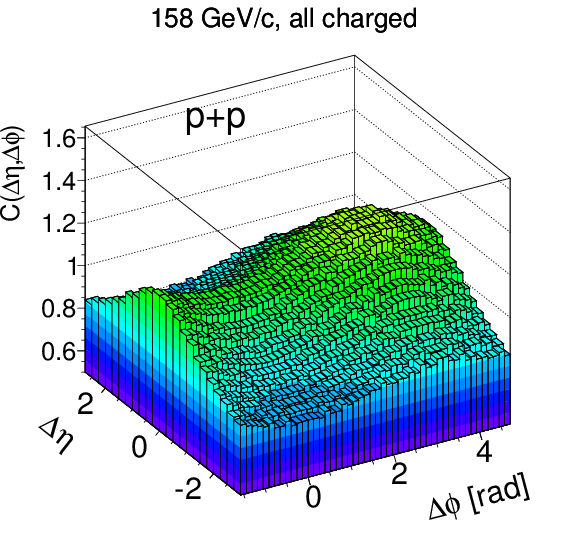}
  \includegraphics[width=0.24\textwidth]{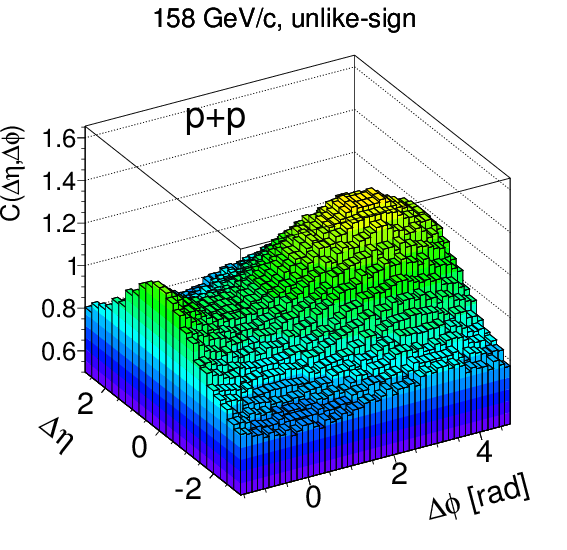}
  \includegraphics[width=0.24\textwidth]{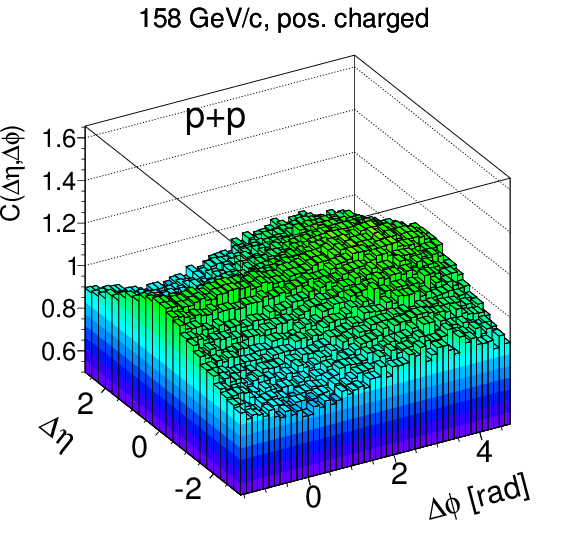}
  \includegraphics[width=0.24\textwidth]{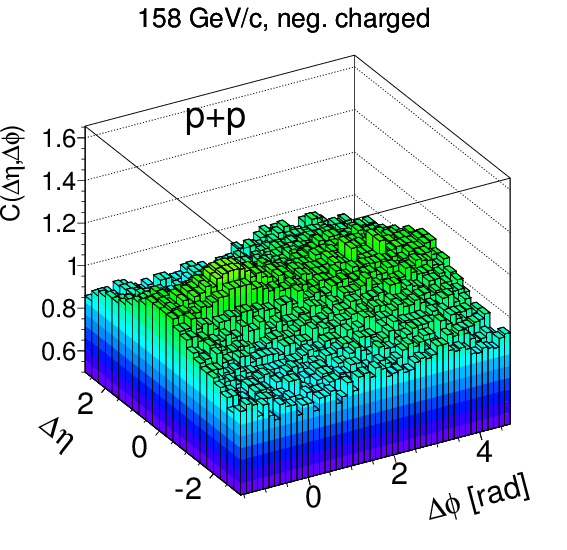}
  \caption{Comparison between Be+Be (top row) and p+p results~\cite{Aduszkiewicz:2016mww} (bottom row) presented for all charge combinations for beam momentum 150$A$ or 158~GeV/c (for Be+Be and p+p, respectively).}
  \label{fig:BeBe_vs_pp}
\end{figure}

 Be+Be correlations were presented so far in vertical scale range $(0.9,1.1)$ while p+p in $(0.5,1.6)$. If Be+Be results were presented in p+p scale (middle plot in Fig.~\ref{fig:scales_comparison}), the correlation structures would be almost flat. In order to allow qualitative comparison between results from both systems, the vertical scales were tuned to present the structures with similar magnitudes. Example comparison for 150$A$/158~GeV/c beam momentum is shown in Fig.~\ref{fig:BeBe_vs_pp}.

Although correlations is Be+Be are generally weaker than in p+p, the structures appearing in both systems are similar and show maximum at away-side and small enhancement at near-side. However, in Be+Be the peak around $(\Delta\eta,\Delta\phi) = (0,0)$ is more prominent with respect to the rest of correlations landscape.

\section{Summary}

For the first time two-particle correlations in Be+Be collisions at SPS beam momenta were presented. The results show a prominent enhancement at $(\Delta\eta,\Delta\phi) = (0,\pi)$, mostly visible in unlike-sign pairs and weaker in like-sign pairs, produced probably due to resonance decays and momentum conservation. Another visible structure is a small maximum at (0,0) appearing in all charge combinations. In unlike-sign pairs of particles it is probably a result of Coulomb attraction, while in like-sign pairs it may come from interplay of Bose-Einstein and Fermi-Dirac effects.

The \Epos model results are in general similar to real data of \NASixtyOne. The only exception is the lack of near-side enhancement, which is due to no short-range correlations (quantum statistics, Coulomb interactions) generated by the model.

Comparing to smaller system of p+p, the correlations in Be+Be collisions are much weaker due to larger combinatorical background diluting the signal. Qualitatively, the structures in both systems are similar with exception of near-side maximum which is more prominent in Be+Be.

\vspace{12pt}
\noindent
{\bf Acknowledgments:} This work was supported by the National Science Centre, Poland grants 2015/19/N/ST2/01689 and 2015/18/M/ST2/00125.

\end{document}